# Ni$^+$ reactions with  aminoacetonitrile, a potential pre-biological precursor of glycine

**Al Mokhtar Lamsabhi\*, Otilia Mó, Manuel Yáñez\***

*Departamento de Química C-9, Facultad de Ciencias, Universidad Autónoma de Madrid, Cantoblanco, 28049-Madrid, Spain.*

**Jean-Claude Guillemin\***

*Sciences Chimiques de Rennes – Ecole Nationale Supérieure de Chimie de Rennes –CNRS – 35700, Rennes France*

**Violette Haldys, Jeanine Tortajada, Jean-Yves Salpin\***

*Laboratoire d'Analyse et Modélisation pour la Biologie et l'Environnement – Université d'Evry Val d'Essonne – CNRS – Bâtiment Maupertuis, Boulevard François Mitterrand, 91025 Evry, France*

Corresponding author:  Jean-Yves Salpin

Tel: 33 1 69 47 76 47   Fax: 33 1 69 47 76 55

e-mail : jean-yves.salpin@univ-evry.fr

Number of pages (including Table and Figures) : 27

**Abstract**

The gas-phase reactions between Ni$^+$($^2$D$_{5/2}$) and aminoacetonitrile, a molecule of pre-biological interest as possible precursor of glycine, have been investigated by means of mass spectrometry techniques. The MIKE spectrum reveals that the adduct ions [NC-CH$_2$-NH$_2$, Ni$^+$] spontaneously decompose by loosing HCN, H$_2$, and H$_2$CNH, the loss of hydrogen cyanide being clearly dominant. The structures and bonding characteristics of the aminoacetonitrile-Ni$^+$ complexes as well as the different stationary points of the corresponding potential energy surface (PES) have been theoretically studied by DFT calculations carried out at B3LYP/6-311G(d,p) level. A cyclic intermediate, in which Ni$^+$ is bisligated to the cyano and the amino group, plays an important role in the unimolecular reactivity of these ions, because it is the precursor for the observed losses of HCN and H$_2$CNH. In all mechanisms associated with the loss of H$_2$, the metal acts as hydrogen carrier favoring the formation of the H$_2$ molecule. The estimated bond dissociation energy of

aminoacetonitrile-Ni$^+$ complexes (291 kJ/mol) is larger than those measured for other nitrogen bases such as pyridine or pyrimidine and only slightly smaller than that of adenine.

**Introduction:**

The reactions between transition metal cations and organic or inorganic molecules have attracted a great deal of attention in the last two decades, because these processes are involved in a significant number of relevant processes in chemistry and biochemistry [1-4]. Transition metals are generally present in the biological media as solvated ions or complexed by different kinds of peptides and proteins. They can also interact with other biomolecules such as nucleic acids including different effects that can vary from the stabilization of the helix to transcription failures.[5-11] Metal cations may play also an important role in astrochemistry. In this case the interactions may take place strictly in the gas-phase or in the surface of dust particles or meteorites, but normally involve small molecules. Among them, aminoacetonitrile presents a particular interest as a precursor of glycine in astrochemical media. As a matter of fact glycine may be formed in meteorites[12] through the interaction of ammonia , formaldehyde and hydrogen cyanide that produce aminoacetonitrile, which is then finally hydrolyzed to yield glycine. This process can be, however, perturbed by the interaction with metal cations, such as Ni$^+$, which as shown in the present study destroy aminoacetonitrile by the loss of HCN. Also, some photochemical studies suggest that glycine can be formed in ices in the interstellar medium or cometary bodies[13-16], as well. The spectroscopic characterization of this compound by means of infrared spectroscopy in Ar matrices was carried out recently.[17] Several theoretical studies at different levels of accuracy on aminoacetonitrile, [17-19] have been also reported in the literature. However, very little is known about the intrinsic reactivity of this compound  and only very recently its gas-phase basicity has been measured by means of ion cyclotron resonance (ICR) techniques.[20] We aim here at characterizing the reactivity of this interesting species with respect to $Ni^+$, a prototype of open-shell transition metal cation with a $3d^9$ ($^2D_{5/2}$) electronic ground state. The interplay between mass spectrometry techniques and density functional theory (DFT) calculations would allow to gain some understanding on the behavior and bonding of aminoacetonitrile-$Ni^+$ that, in principle, may be useful to rationalize the behavior of more complicated systems which present similar basic sites.

**Experimental**

***Synthesis*** Aminoacetonitrile has been prepared as recently reported.[20]

***Mass spectrometry:***

All experiments were carried out using a VG Analytical ZAB-HSQ hybrid mass spectrometer of BEqQ geometry which has been described in detail previously.[21] Complexes were generated by the CI-FAB method.[22-27] The CI-FAB source was constructed from VG Analytical EI/CI and FAB ion source parts with the same modifications described by Freas et al.[22] The conventional FAB probe tip has been replaced by a nickel foil of high purity. "Naked" metal ions were generated by bombardment with fast xenon atoms (Xe gas 7-8 keV kinetic energy, 1-2 mA of emission current in the FAB gun). The organic samples were introduced via a probe in a non-heated source. We can assume that due to the relatively high pressure in that source ($10^2$-$10^3$ Pa), efficient collisional cooling of the generated ions takes place. Therefore we will consider that excited states of the $Ni^+$ ions which could be formed in these experimental conditions are not likely to participate in the observed reactivity as already postulated by Hornung et al.[26] The ion beam of the $Ni^+$ adduct complexes formed with aminoacetonitrile were mass-selected (using an acceleration voltage of 8 kV) with the magnetic analyser B. The ionic products of unimolecular fragmentations, occurring in the second field-free ($2^{nd}$ FFR) region following the magnet, were analyzed by means of Mass-analyzed Ion Kinetic Energy MIKE[28,29] by scanning the electric sector E.

The CAD (Collision Activated Dissociation) experiments were carried out in the same fashion but introducing Argon in the cell as the collision gas. The pressure of argon in the collision cell was adjusted so that the main beam signal was reduced by approximately 30%. The spectra were recorded at a resolving power (R) of ~1000

**Computational details:**

All quantum chemistry calculations presented in this paper have been carried out with the B3LYP hybrid DFT method[30,31] as implemented in the Gaussian03 series of programs.[32] The geometries of the different species under consideration were optimized using the all-electron basis (14s9p5d/9s5p3d) of Wachters[33] and Hay[34] augmented by a set of $f$ functions for Ni and the 6-311G** basis set for remaining atoms of the system. The same basis set expansion and method were used to calculate the harmonic vibrational frequencies, in order to classify the stationary points of the potential energy surface (PES) as local minima or transition states, and to evaluate the corresponding zero-point energies (ZPE), which were scaled by the empirical factor 0.9806.[35] In all the cases, the $<S^2>$ expectation value showed that the spin contamination of the unrestricted wave function was always very small.

The bonding characteristics, as well as the electron density redistributions triggered by $Ni^+$ association were analyzed by means of the atoms in molecules (AIM) theory.[36] For this purpose we have located the relevant bond critical points and evaluated the charge density at each of them. To perform the AIM analysis we have used the AIMPAC series of programs.[37] Also a second order perturbation method in the framework of the natural bond orbital (NBO) approach[38] was used to evaluate the interactions between orbitals of the base and orbitals of the metal, involved in the dative bonds from the former to the latter and possible back donations from the latter to the former. Since all complexes are open-shell systems, the NBO analysis has to be carried out for both the α- and β-sets of MOs. However, for the sake of simplicity, hereafter we will provide only the information corresponding to orbital interactions within the β-subset, because this is the subset that contains the *3d* as well as the *4s* unoccupied orbitals. The $\alpha$ natural orbital set exhibits a similar behavior although in this case only the empty *4s* orbital is within the subset.

**Results and discussion:**

*Mass spectra*

The $^{58}Ni^+$ ions react with neutral aminoacetonitrile to produce $[^{58}Ni\text{-}NCCH_2NH_2]^+$ adduct ions at m/z 114. The unimolecular decomposition of the $[^{58}Ni\text{-}NCCH_2NH_2]^+$ complex have been investigated by means of MIKE analysis to obtain information related to the structure and reactivity of this complex. The MIKE spectrum is shown in Figure1. The [Ni-NCCH_2NH_2]^+ ion undergoes fragmentation according to several disscociation pathways. The main fragmentation corresponds to loss of [H,C,N] to produce [Ni-C,H_3,N]^+ ion at m/z 87, the base peak of MIKE spectrum. This reactivity differs from that observed with alkanenitriles, for which loss of 27 daltons is not observed. As a matter of fact, previous studies have demonstrated that the unimolecular reactivity of Ni^+/alkanenitrile complexes is characterized by loss of the intact ligand, and beginning with n-propyl cyanide, losses of $H_2$ and alkenes. [39-42] A dramatic increase of the two latter processes with the chain length has been noted. On the other hand, elimination of hydrogen cyanide has been reported for $\alpha$- and $\beta$- unsaturated alkenenitriles.[43] A second peak associated with loss of $H_2$ is observed at m/z 112. Such a dehydrogenation process has been already observed for both alkanenitriles [39-41] and primary amines. [44-46] but its intensity is presently rather weak compared to what has been reported for primary amines. Another significant difference between the unimolecular reactivity of Ni^+/aminoacetonitrile and Ni^+/amine systems is the absence of elimination of nickel hydride NiH in the former case. To complete this survey, another two small peaks are also observed at m/z 113 and m/z 85. They correspond to the elimination of H$\cdot$ and [C,H_3,N], respectively.

Ions at m/z 85 have been already observed in significant intensity on metastable spectra Ni$^+$/alkenenitriles adducts [43], or more recently on the CID spectrum of electrospray-generated [Ni(NC-CH$_2$-N$_3$), -N$_2$]$^+$ complexes.[47] Finally, it is worth noting that bare Ni$^+$ cation is not observed under metastable conditions while it is the predominant process for short-length alkylnitriles. [41]

Under CAD conditions one can note in the resulting CAD spectrum displayed in Figure 2, that the intensity of the peak corresponding to the loss of [H,C,N] increases showing that this decomposing channel is favored when energy is provided. Formation of m/z 85 and 87 ions could correspond to elimination of methanimine and hydrogen cyanide, respectively. Since the ability of transition metal ions to be dicoordinated is well known, the competitive losses of neutral HCN and CH$_2$NH might be probably arise from Ni$^+$-bound heterodimers such as [HCN-Ni$^+$-NH=CH$_2$], which may undergo competitive dissociations leading to the two fragment ions at m/z 85 and m/z 87.

We can also observe in this spectrum the presence of two minor ions at m/z 58 and 60. The first one corresponds to bare $^{58}$Ni$^+$ generated by the elimination of the intact ligand. But, this process occurs at a very minor extent compared to monofunctional molecules such as nitriles and primary amines. The second one might be H$_2$Ni$^+$. However, formation of this particular ion has never been observed in previous studies, neither with amines nor with nitriles. This ion at m/z 60 could also correspond, at least to some extent, to $^{60}$Ni$^+$. As a matter of fact, the observation of dehydrogenation in the metastable spectrum of the [$^{58}$Ni-NCCH$_2$NH$_2$]$^+$ complex strongly suggests that the $^{58}$Ni-complex generated in the FAB source may interfere with the corresponding dehydrogenation product of the $^{60}$Ni complex, namely [$^{60}$Ni-NCCH$_2$NH$_2$, -H$_2$]$^+$ , which in turn could give rise to bare $^{60}$Ni$^+$ ion. Additional experiments such as MIKE and CAD spectrum of both m/z 112 and [$^{60}$Ni-NCCH$_2$NH$_2$]$^+$ complex (m/z 116) could have raised the uncertainty about the structure of the m/z 60 ion, but could not be performed because of failure of our FAB instrument. However, complementary electrospray experiments performed by Dr. D. Schröder confirmed that $NiH_2^+$ species is indeed generated. These spectra were recorded using a VG BIO-Q tandem QHQ mass spectrometer (Q stands for quadrupole and H for hexapole), which has been described elsewhere.[48] 1 mg of $NiI_2$ was dissolved in 1 ml distilled water, then 50 µl of a 5% solution of freshly made $H_2NCH_2CN$ in distilled water was added and the solution was measured, thus allowing ions of the type $(H_2NCH_2CN)NiI^+$ to be generated. By adopting harsh ionization conditions, in-source loss of atomic iodine is observed, thereby allowing reduction of Ni(II) to Ni(I)[49] and formation of $[Ni-NCCH_2NH_2]^+$ complex. MS/MS spectra of this ion with either argon or xenon as collision gas exhibit a *m/z* 60 at elevated collision energy, undoubtedly attributed to $NiH_2^+$, as hydrogen loss(es) is not observed with the QHQ intrument. Note that in sector experiments, hydrogen losses are often much preferred in detection.[50] Finally, one may assume that $Ni^+$-bound heterodimer such as $[H_2-Ni^+-NH=CH.CN]$ could also be the precursor of both m/z 112 and m/z 60 ions.

In order to assess the mechanisms behind the aforementioned experimental findings we have carried out a detailed study of the [aminoacetonitrile-$Ni^+$] potential energy surface by means of DFT calculations.

*Coordination of Nickel$^+$*

The optimized geometries of the different conformers of $Ni^+$-$NCCH_2NH_2$ and a selection of some structural parameters are shown in Figure 3. Total energies and $Ni^+$ binding energies of all the structures (minima, transitions states and fragments) considered in this study are summarized in Table S1 of the supporting information.

Previous studies about the reactivity of transition-metal ions (and notably $Ni^+$) with nitriles have suggested the co-existence of both "end-on" and "side-on" coordination modes.[40,41,51] The first one corresponds to the interaction of the metallic center with the lone pair of the nitrogen of the cyano group in a linear M···N≡C arrangement (for monofunctional ligands) while the second one involves interaction with the π-system. Accordingly, two possible types of coordination have been considered: $π$-type interactions with the C≡N group which yields structures **a** and **d** (side-on), and σ-type interactions with the nitrogen lone pair of either the cyano or the amino groups, which leads to structures **b** and **c**, respectively (end-on). The most stable conformer has been found to be the complex side-on **a**, which is stabilized through the interaction of the metal cation with the nitrogen lone pair of the amino group and also through a $π$-type interaction with the C≡N group. As a matter of fact, a second order NBO analysis shows both interactions to be rather strong, even though the former is stronger than the latter. As illustrated in Table 1, for complex **a** there are two empty *sd*-type hybrid orbitals on the metal which participate in the interaction with aminoacetonitrile. The first of these interactions corresponds to a dative bond from one of the C-N π-bonding orbitals toward the second empty hybrid of Ni, with a 53 % participation of the *4s* orbital. The second one is another dative bond from the lone pair of the amine group nitrogen into the first empty hybrid on Ni, with a 39% *4s* character. Both dative interactions are followed by a backdonation from *3d* occupied orbitals of Ni towards $π_{CN}*$ antibonding orbital. The charge transfer from the $π_{CN}$ bonding and the population of the $π_{CN}*$ antibonding orbital is reflected in the lengthening of the bond (by 0.014 Å), as well as in a decrease of the charge density at the corresponding bond critical point (see Figure 4). The involvement of the amino lone-pair in the interaction with the metal cation results also in a significant lengthening (0.048 Å) of the $C-NH_2$ bond and in a concomitant decrease of the charge density at the corresponding bond critical point. It is worth noting that, coherently with the NBO picture, the molecular graph of complex **a** shows two bond paths with origin in $Ni^+$, connecting the metal to the amino nitrogen and to the CN group, as well as the presence of a ring critical point.

The second less stable complex **b**, which lies 18 kJ/mol above the global minimum **a**, corresponds to the end-on interaction of $Ni^+$ with the cyano nitrogen lone pair. The NBO analysis shows the existence of a very strong interaction between this lone pair and a *sd* hybrid on Ni with a large *s* character (77%), followed by a backdonation from an occupied *d* orbital of Ni towards a $\pi_{CN}$* antibonding orbital, which results in a slight lengthening of the CN bond (see Figure 3). The less stable complex **d** is that in which $Ni^+$ only interacts with the CN group. The difference in energy with the complex **a** (70 kJ $mol^{-1}$) gives a qualitative estimate of the extra-stabilization provided by the amino group. The NBO analysis indicates that both $\pi$- and $\sigma$-interactions take place (see Table 1) although the former are clearly dominant. As in the other complexes a backdonation from occupied *3d* orbitals of the metal cation into the $\pi_{CN}$* antibonding orbital also occurs. These interactions and the subsequent polarization of the C-C and the $C-NH_2$ bonds result in a lengthening of both the CN and the CC bonds by 0.03 Å and 0.037 Å, respectively and in a shortening of the $C-NH_2$ linkage by 0.032 Å. The attachment of $Ni^+$ to the amino group leads to structure **c** which is 33 kJ/mol less stable than the global minimum. In this case, only a dative bond from the amino nitrogen lone pair toward an empty *sd* hybrid on Ni, with a 53% of *s* character is found. As expected this involves a lengthening of the $C-NH_2$ bond by 0.048 Å and a concomitant shortening of the C-C bond by 0.018 Å.

*Reactivity of the $[Ni-NCCH_2NH_2]^+$ adducts*

As mentioned above, the MIKE and MIKE-CAD spectra of $[^{58}Ni-NCCH_2NH_2]^+$ shows different fragmentation processes, that we may attribute to loss of $H_2$, HCN or $CH_2NH$. The loss of HCN and $H_2$ being the dominant ones, we will concentrate our attention in these two processes in particular.

As we shall discuss shortly, the different mechanisms associated with the two dominant unimolecular processes may have its origin not only in the global minimum but also in the other less stable aminoacetonitrile-Ni$^+$ complexes. This is consistent with previous experimental studies dealing with the reactivity towards nitriles, which suggested that a single type of interaction could not account for all the fragmentration processes observed.[40,43] However, as shown in Figure 5 all of them are connected through rather small activation barriers, but more importantly, well below the entrance channel. The most stable structure **a** may evolve to complex **b** by a simple rotation of Ni in the plane molecule, through a barrier of 54 kJ/mol. A rotation of the metal cation out of the plane of the molecule would connect **a** and **d**. The evolution from **a** to **c**, through an energy barrier of 45 kJ/mol, implies an internal rotation of the –NH$_2$Ni group.

*HCN-Loss*

The observed loss of HCN dominates both the MIKE and CAD spectra and requires unavoidably a hydrogen shift towards the carbon atom of the CN group. This is possible from either the global minimum **a** or from the local minimum **b**. Both processes correspond to H-transfer from the methylene group toward the carbon atom of the CN group leading to the cyclic minimum **2** (see Figure 6), where nickel is bisligated to the two nitrogen atoms. However, the process with origin in the global minimum **a** should be discarded because it involves an activation barrier which is not only above the entrance channel (See Figure 6), but also much higher than the barrier associated with the **a** → **b** isomerization.

The most favorable process from the cyclic minimum **2** is an $\alpha$-C-C bond insertion of the metal yielding minimum **3**, which may eventually dissociate into HCN + [NiCH=NH$_2$]$^+$ or alternatively into [NiNCH]$^+$ + HC-NH$_2$, the former process being much more favorable from the thermodynamic point of view than the latter, which is in agreement with the experimental finding that the dominant loss corresponds to HCN and not to [C,H₃,N]. One may also consider the possibility of inverting the order of the processes, by inserting first the metal into the C-C bond and having the hydrogen shift as a second step. The global minimum **a** is a good starting point for such a mechanism due to the bridging position of the metal between the two basic sites of the aminoacetonitrile. In fact the insertion leads to the local minimum **1** through the transition state **TSa_1**. It can be noticed that the insertion is accompanied by an internal rotation of the C≡N group, because as soon as the CC bond cleaves, the interaction of the nitrogen atom of the CN fragment with the metal cation is very strong, reflecting its high intrinsic basicity. From minimum **1**, two hydrogen shifts can be envisaged. The most favorable one, involving the **TS1_3** transition state, involves a methylene hydrogen and leads also to structure **3**, already found in the first mechanism discussed above with origin in the cyclic minimum **2**. The second possible hydrogen shift involves a amino hydrogen and leads to a very stable local minimum **4**, which lies 172 kJ mol$^{-1}$ below the most stable aminoacetonitrile-Ni$^{+}$ adduct, **a**. As shown in Figure 6, minimum **4** could also be formed by a hydrogen shift from **3**. Similarly to minimum **3**, structure **4** would eventually dissociate either by losing HCN or [C,H₃,N]. There is a difference however between both mechanisms as far as the loss of HCN is concerned. When dissociation originates in minimum **4**, in the accompanying ion product, Ni$^{+}$ is attached to the imino nitrogen of H₂C=NH, while when the dissociation originates in minimum **3** it is attached to the carbon atom of HC-NH₂. In summary, the observed loss of HCN can be associated with two different mechanisms with origin in adducts **a** and **b**, respectively, that lead to the same precursor **3** and which involved rather similar activation barriers. The alternative mechanisms for the loss of HCN have as precursor structure **4**, but the activation barriers to reach this minimum are much higher than those to yield **3**, and therefore we may conclude that the majority of the [Ni, H$_3$, C, N]$^+$ product ions will be in the form of [NiCH=NH$_2$]$^+$ complexes.

It is worth noting also that the loss of [C,H$_3$,N] in the form of CH=NH$_2$ is slightly endothermic, while the loss of CH$_2$=NH is highly exothermic. Hence, very likely the observed [C,H$_3$,N] loss comes mostly from the fragmentation of structure **4**. It is also important to emphasize that according to our theoretical survey the [H,C,N,Ni]$^+$ cation has always a HCNNi$^+$ connectivity. Actually an ion with an H-Ni-CN structure lies much higher in energy.

Some other mechanisms with origin in the local minimum **2** lead also to the loss of HCN (or the loss of H$_2$), but they involve larger activation barriers than those discussed above (see Figure 1S of the supporting information).

Some mechanisms leading to the loss of HNC instead of HCN have also been investigated. However, all hydrogen transfers either from the amino group or from the CH$_2$ toward the cyano group, involved activation barriers that were well above the entrance channel (51 and 101 kJ mol$^{-1}$, respectively). Since these mechanisms are not likely to take place they have not been included in Figure 6.

*H$_2$-Loss*

Many possibilities may be considered for the loss of H$_2$. In our survey we have analyzed a large number of processes, but for the sake of conciseness we are going to summarize the most favorable ones. As we shall show along this section, starting from complexes **c** or **a,** the exit channels are much lower in energy than the one mentioned above (see Figure 7). All these mechanisms have in common that the nickel atom acts as hydrogen carrier favoring the formation of the H$_2$ molecule. In fact, starting from **c**, two pathways are opened, a hydrogen transfer from the amino group to nickel (through **TSc_10**) followed by a second H-transfer from the methylene group (through **TS10-12**), or a process where these two steps are inverted, i.e., the first step is a H-transfer from the methylene group (through **TSc_8**) toward Ni, and the second one a H-transfer form the amino group (**TS8_12**). In both cases the same local minimum **12** is reached. As illustrated in Figure 7, the last possibility is clearly more favorable. Even though **TS8_12** is slightly above the entrance channel by 26 kJ mol$^{-1}$, this process can be observed because under the experimental conditions the local minima may have enough internal energy to overpass barriers slightly higher than the entrance channel. On the other hand, since the process corresponds to a H-transfer the existence of tunneling effects should not be discarded. Both situations would be consistent with the fact that the $H_2$ loss is observed in much less proportion than the loss of HCN. Finally, note that we also considered a multi-center transition state (MCTS) for the **9 → 13** interconversion, since MCTS usually provide rather low-lying pathways. [52,53] However, it turned out that optimization of such a MCTS failed, because one of the hydrogen atom moves during the optimization process and the initial structure finally collapses back to the intermediate **9**.

The most stable adduct **a** may be also a precursor for the loss of $H_2$, but as shown in Figure 7, the mechanism through the intermediates **9**, **11** to reach **13**, involves a quite high activation barrier associated with the **9 → 11** isomerization, and therefore should not compete with the mechanism with origin in adduct **c** and preliminary **a → c** interconversion.

As a final remark, it is also worth noting that the estimated bond dissociation energy of aminoacetonitrile-Ni$^+$ complexes (291 kJ/mol) is larger than those measured for other nitrogen bases such as pyridine (256 ± 15)[9] or pyrimidine (244 ± 9)[54] and only slightly smaller than that of adenine (297 ± 10)[55], in which the metal ion also forms a chelate structure bridging between the N7 position of the imidazolic ring and the amino group.

**Conclusions**

The MIKE spectra of [NC-CH$_2$-NH$_2$, Ni$^+$] ions show that they spontaneously decompose by loosing HCN, H$_2$, and [H$_3$,C,N], the loss of hydrogen cyanide being clearly dominant. A survey of the corresponding potential energy surface, shows that the direct insertion of the metal cation into the C-C bond of the most stable adduct or the insertion through the cyclic intermediate **1**, in which Ni$^+$ is bisligated to the cyano and the amino group are the most favourable mechanisms leading to the observed losses of HCN and [H$_3$,C,N]. The loss of CH=NH$_2$ is predicted to be endothermic, while the loss of CH$_2$=NH is clearly exothermic, so that we can safely conclude that the observed loss of [H$_3$,C,N] corresponds mostly to CH$_2$=NH. Our theoretical calculations also indicate that in the [H,C,N,Ni]$^+$ accompanying ion, the metal is attached to the N atom of the HCN molecule. In all the mechanisms associated with the loss of H$_2$, the metal acts as hydrogen carrier favoring the formation of the H$_2$ molecule. The topology of the PES also shows that the *m/z* 87 peak observed in both the MIKE and the CAD spectra corresponds exclusively to the loss of HCN, because the mechanisms associated with the loss of HNC involve activation barriers much higher in energy than the entrance channel.

**Acknowledgement:** This work has been partially supported by the DGI Project No. CTQ2006-08558/BQU and by the Project MADRISOLAR, Ref.: S-0505/PPQ/0225 of the Comunidad Autónoma de Madrid. Authors would like to warmly acknowledge Professor Detlef Schröder for kindly performing complementary experiments. A.M.L gratefully acknowledges a Juan de la Cierva contract from the Ministerio de Educación y Ciencia of Spain. We also acknowledge a generous allocation of computer time at the Centro de Computación Científica de la Universidad Autónoma de Madrid. J.-C. G. thanks the GDR CNRS "Exobiologie" and the Centre National d'Etudes Spatiales for financial support.

**Supporting information** Total, zero-point energies (ZPE) and relative energies of the various structures considered are provided as supporting information, together with potential energy surfaces associated with alternate mechanisms leading to the ions observed experimentally.

**Table 1.** NBO analysis of aminoacetonitrile-Ni$^+$ adducts showing the energies[a] (kJ mol$^{-1}$) associated with dative bonds from $\pi_{CN}$ bonding orbitals of the cyano group and from the nitrogen atoms lone pairs toward the empty orbitals of the metal. The last two rows report the percentage of the $s$ and $d$ character of the Ni$^+$ empty hybrid orbital. This analysis corresponds to $\pi$ spin-orbitals.

| Complex | **a** | **b** | **c** | **d** |
|---|---|---|---|---|
| $\pi_{C\equiv N} \rightarrow$ LP* (1) (Ni) | - | - | - | 54.1 |
| $\pi_{C\equiv N} \rightarrow$ LP* (2) (Ni) | 61.9 | - | - | 18.1 |
| $n_{N(\equiv C)} \rightarrow$ LP* (1) (Ni) | - | 125.4 | - | 23.8 |
| $n_{N(\equiv C)} \rightarrow$ LP* (2) (Ni) | - | - | - | - |
| $n_N \rightarrow$ LP* (1) (Ni) | 87.5 | - | 97.0 | - |
| $n_N \rightarrow$ LP* (2) (Ni) | - | - | - | - |
| $nd_{Ni} \rightarrow \pi^*_{C\equiv N}$ | 69.0 | 23.0 | - | 62.2 |
| LP* (1) Ni | 39% $s$, 61% $d$ | 77% $s$, 23% $d$ | 53% $s$, 47% $d$ | 93% $s$, 7% $d$ |
| LP* (2) Ni | 53 % $s$, 47% $d$ | 97% $s$, 3% $d$ | 42 % $s$, 58 % $d$ | 1% $s$, 99% $d$ |

[a] Only interaction energies larger than 8 kJ mol$^{-1}$ are included

**Figure Captions**

**1.** MIKE spectrum of $[^{58}Ni\text{-}NCCH_2NH_2]^+$ ions. Mass to charge ratios of the fragment ions are deduced from the ratio of their kinetic energy to that of the precursor ion (m/z 114).

**2.** CAD spectrum of $[^{58}Ni\text{-}NCCH_2NH_2]^+$ ions. Mass to charge ratios of the fragment ions are deduced from the ratio of their kinetic energy to that of the precursor ion (m/z 114).

**3.** Optimized geometries of the aminoacetonitrile-$Ni^+$ adducts. Bond lengths are in Å and bond angles in degrees.

**4.** Molecular graphs of the aminoacetonitrile-$Ni^+$ adducts. Red dots represent bond critical points and yellow dots ring critical points. Electron densities are in a.u.

**5.** Isomerization barriers between the most stable aminoacetonitrile-$Ni^+$ adduct **a** and the other aminoacetonitrile-$Ni^+$ adducts. Relative energies are in kJ $mol^{-1}$.

**6.** Energy profile associated with the different mechanisms that lead to HCN and $H_2CNH$ losses in aminoacetonitrile-$Ni^+$ gas-phase reactions. Energies relative to the most stable aminoacetonitrile-$Ni^+$ adduct **a** are in kJ $mol^{-1}$.

**7.** Energy profile associated with the loss of $H_2$ in aminoacetonitrile-$Ni^+$ gas-phase reactions. Energies relative to the most stable aminoacetonitrile-$Ni^+$ adduct **a** are in kJ $mol^{-1}$.

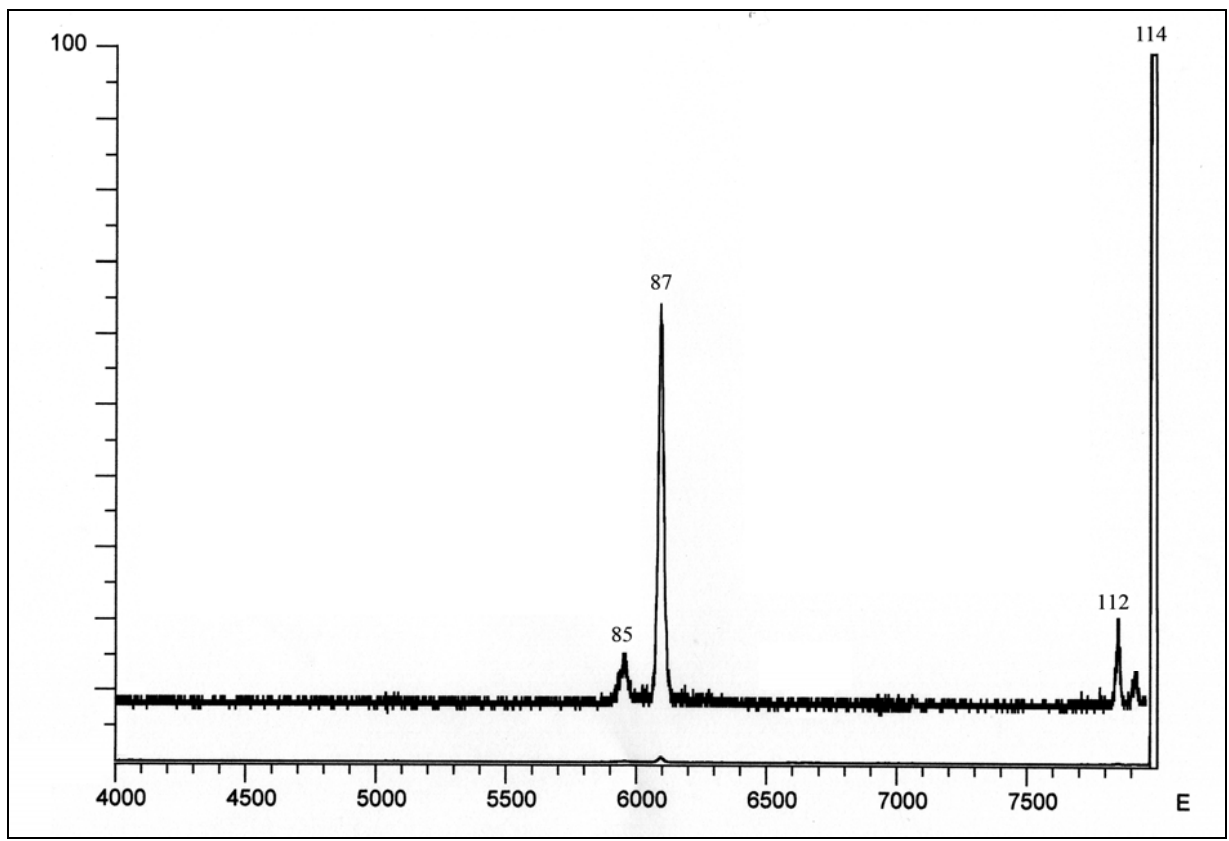

**Figure 1**

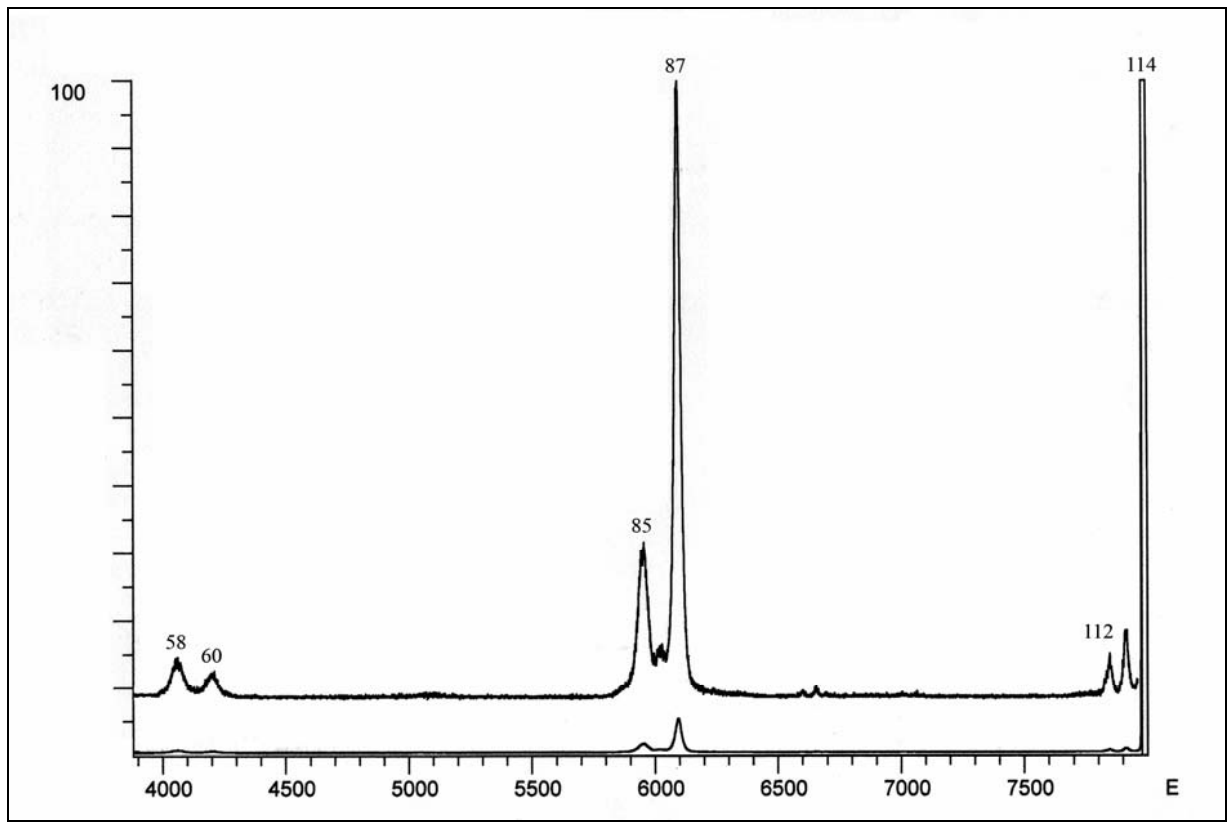

**Figure 2**

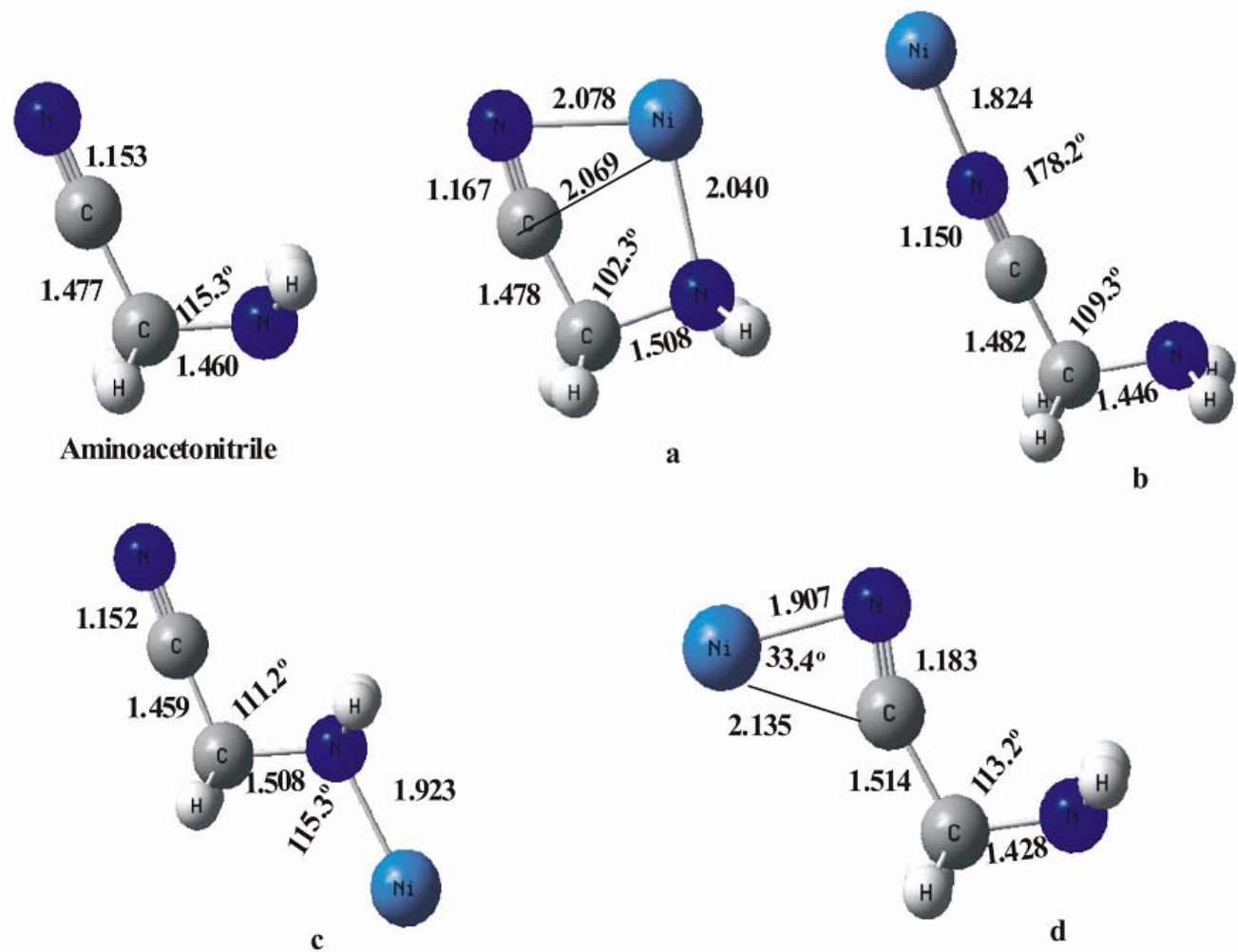

**Figure 3**

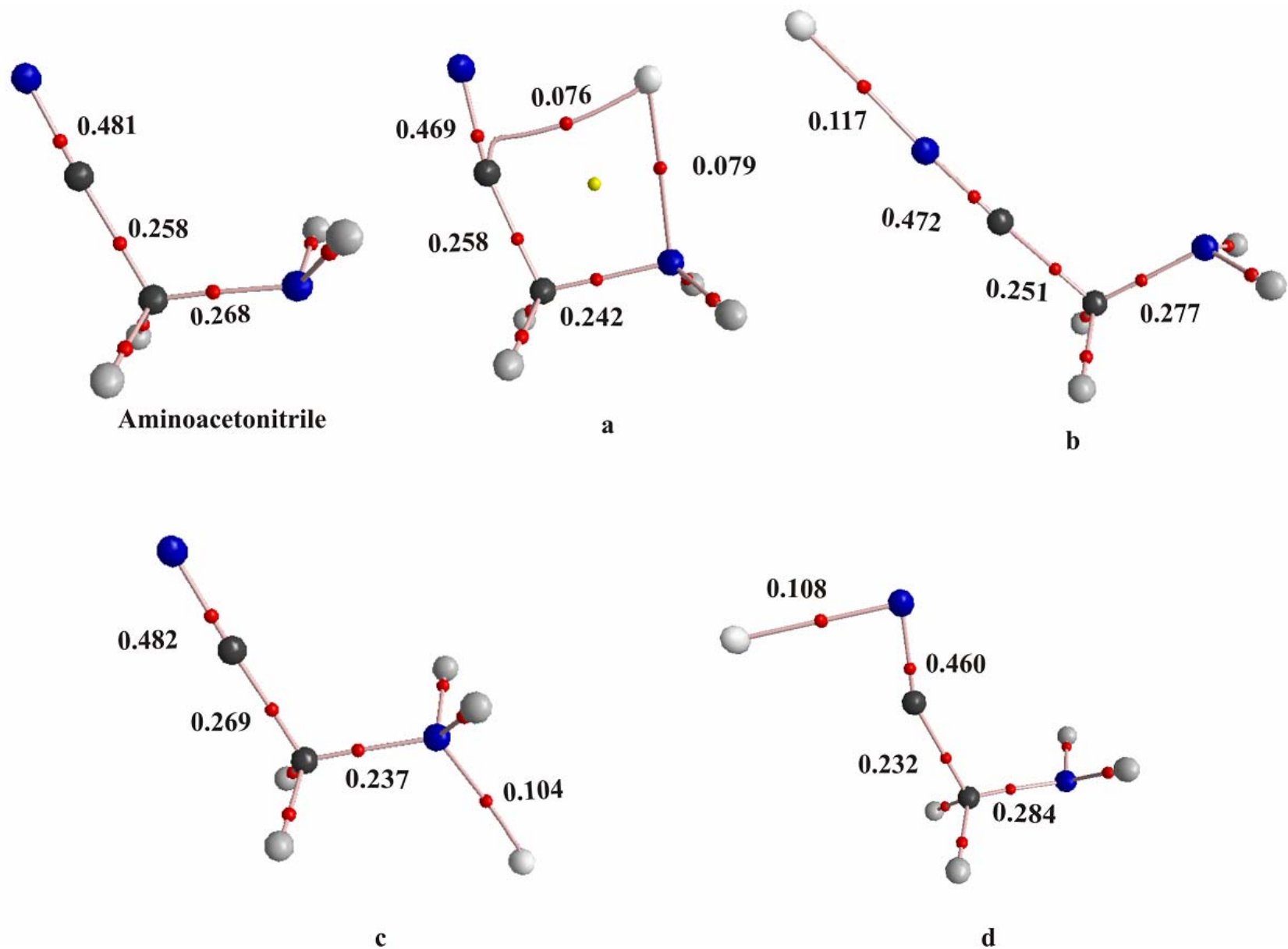

**Figure 4**

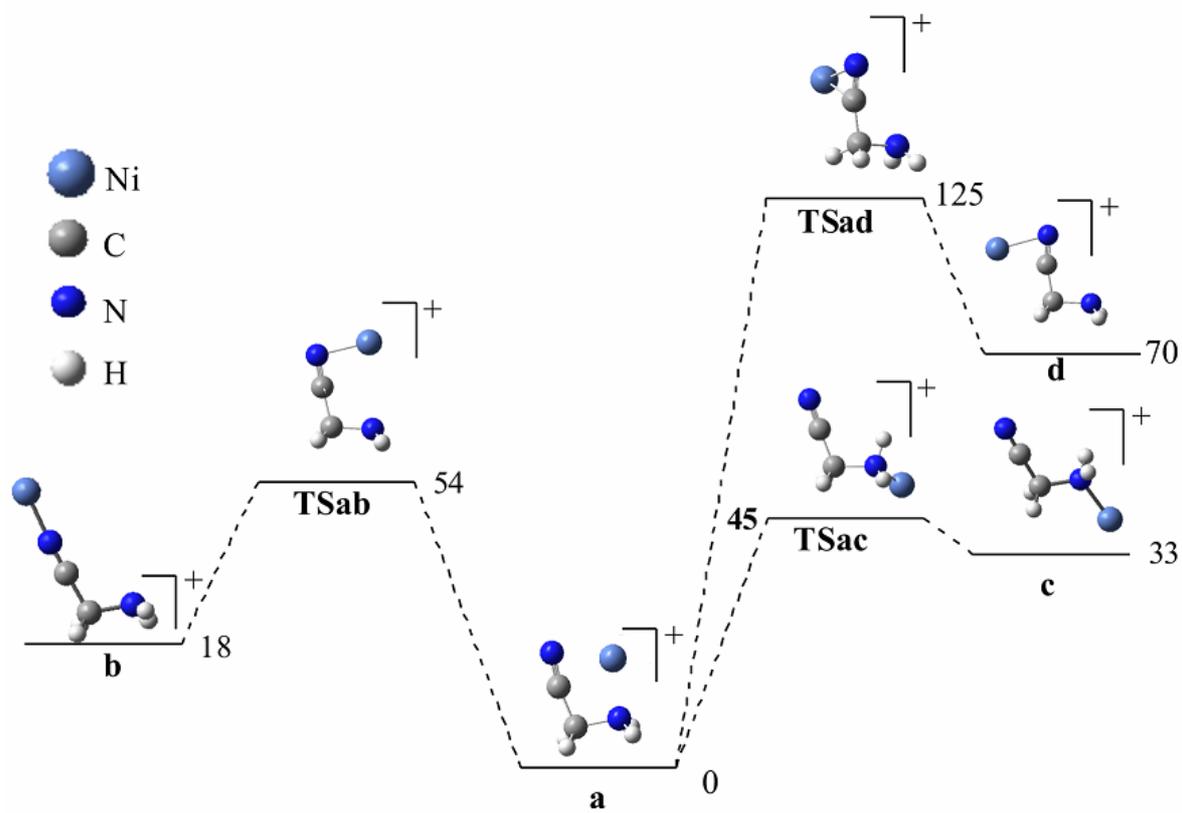

**Figure 5**

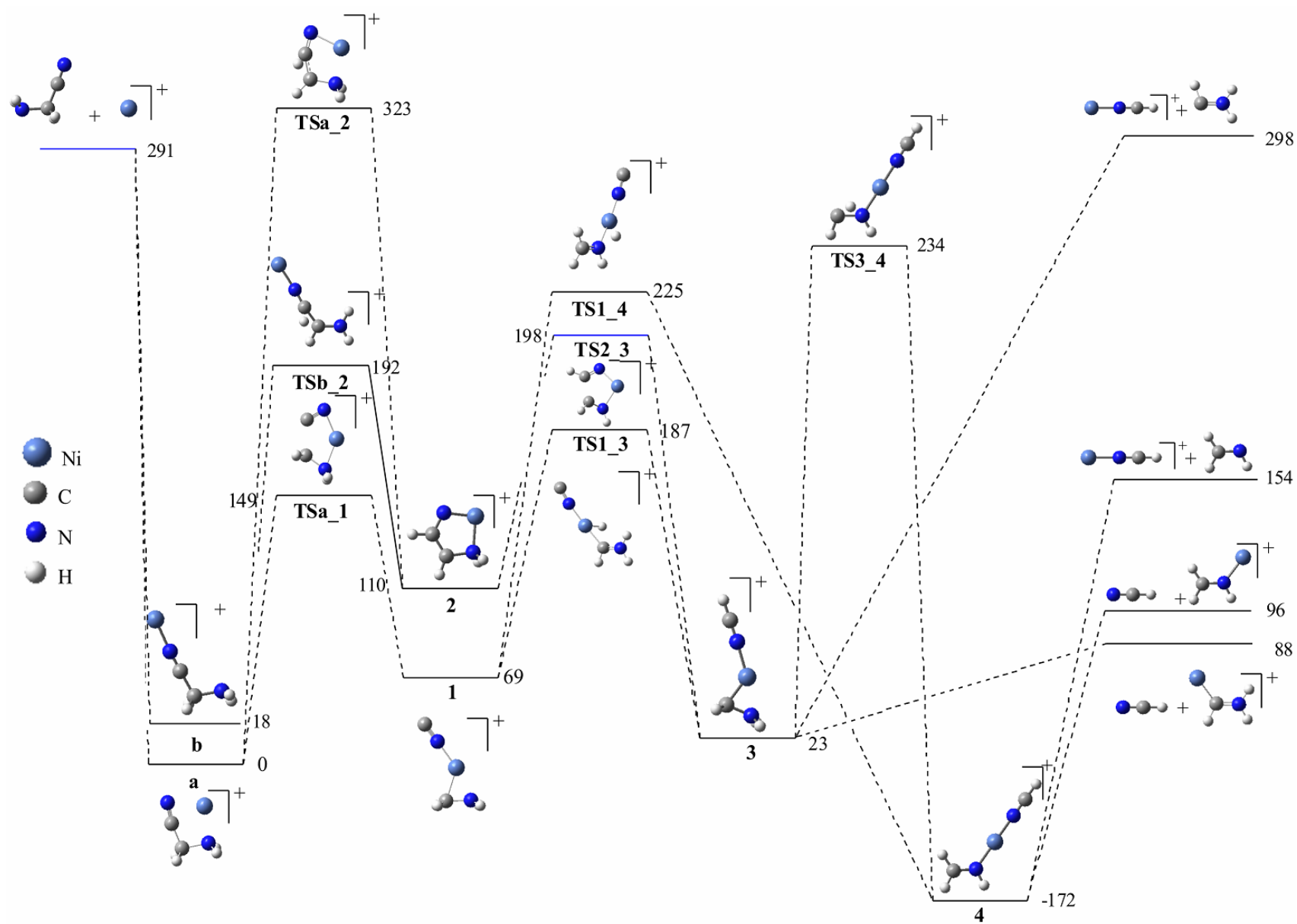

**Figure 6**

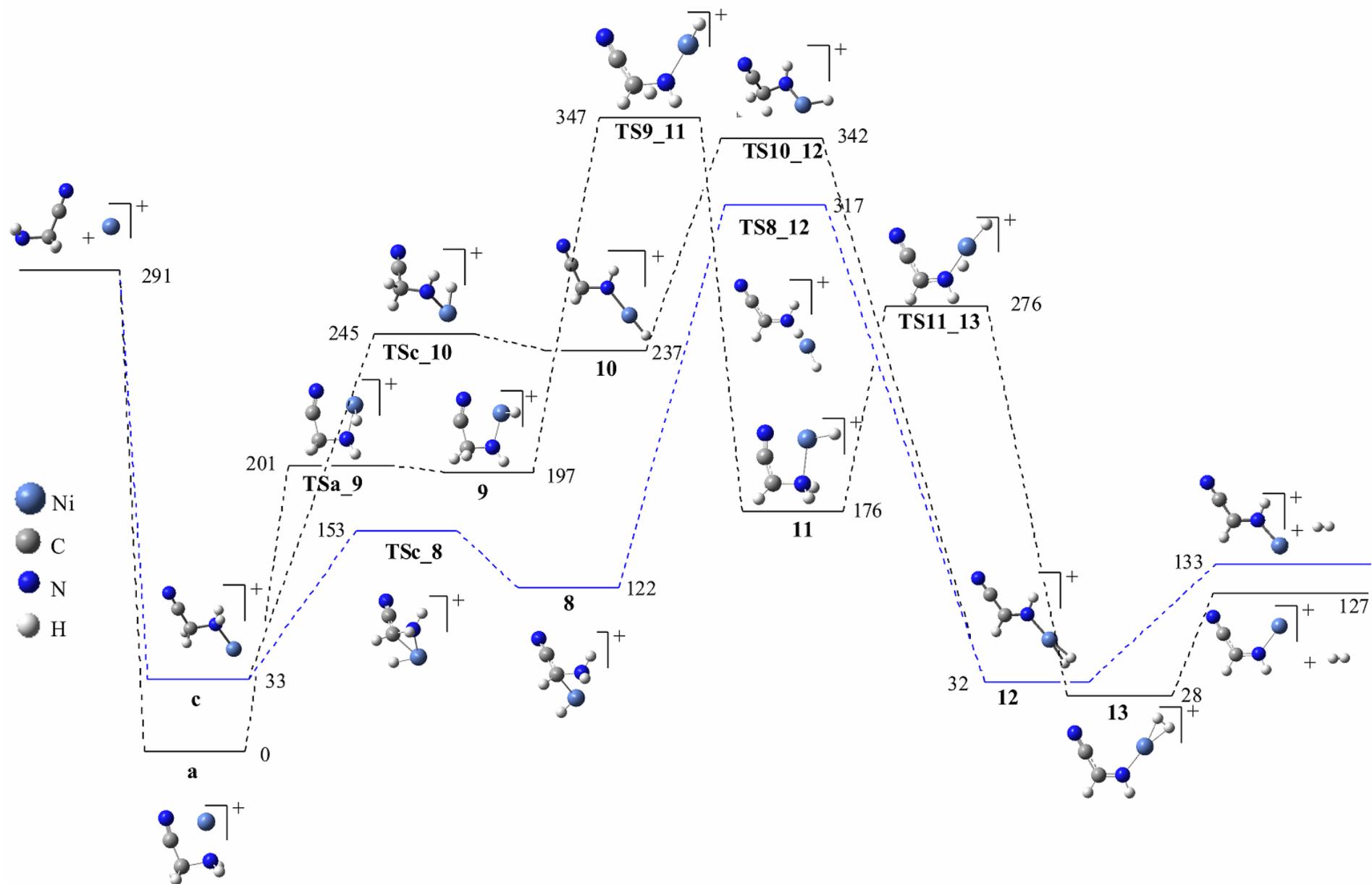

**Figure 7**